\documentstyle[12pt]{article}
\begin{document}
\setlength{\baselineskip}{5.9mm}

\title{Isospin breaking in the reaction $np\rightarrow d\pi^0$
at threshold}
\author{J. A. Niskanen\\ Department of Physics, P.O. Box 9, \\
FIN--00014 University of Helsinki, Finland}
\date{  }
\maketitle
\begin{abstract}
The model for charge symmetry breaking in the reaction
$np\rightarrow d\pi^0$ applied earlier around the $\Delta$
region is used to calculate the integrated forward-backward
asymmetry of the cross section close to threshold. The
mixing of the $\pi$ and $\eta$ mesons appears as strongly
dominant at these energies. This contrasts elastic $np$
scattering experiments, where the $np$ mass difference in
OPE dominates, or $np\rightarrow d\pi^0$ closer to the
$\Delta$ region.
\end{abstract}

\section{Introduction}
Tests of charge symmetry and its breaking (CSB) in elastic
two nucleon interaction fall into two categories according
to how the investigated interaction behaves with respect to
the total isospin.
In the classification
of Henley and Miller \cite{hm} a class III force depends on the
zeroth component (in spherical tensor notation)
of the total isospin as $\tau_{10}+\tau_{20}$
and is nonzero only for the $pp$ and $nn$ states (with opposite
signs for the two). This has been investigated for decades
in low-energy $NN$ scattering and mirror nuclei 
with the main difficulties being extraction of 
the Coulomb force in the $pp$ system and the lack of neutron 
targets in $nn$ scattering \cite{mns,wim}. Clearly this
interaction acts only in isospin one states and cannot change 
the value of the isospin. The same is true also for the
isotensor force, class II. In contrast, a class IV force proportional
to either 
$\tau_{10}-\tau_{20}$ or $(\vec\tau_{1}\times\vec\tau_{2})_0$
necessarily changes the isospin and therefore can act only in 
the $np$ system, where both isospin zero and one are allowed.

The origin of class IV forces has three main sources,
which are overall nearly equally important in elastic 
scattering:\footnote{Experiments are performed at a single
angle where the $np$-mass difference in OPE dominates.}
i) the $np$-mass difference, ii) $\rho^0\omega$-meson mixing
and iii) the magnetic interaction of the neutron with the
proton current.
 At the turn of the decade this isospin breaking interaction 
was seen in experiments as a difference of the neutron and
proton analyzing powers $\Delta A = A_n - A_p$
in polarized $np$ scattering \cite{abegg,knutson,greeniaus}.

The class IV force can also show up in an interesting way in
inelasticities. Namely, isospin respecting mechanisms 
in $NN \rightarrow d\pi$ involve only isospin one initial
states. This
sets strict constraints to the spins and parities of the 
initial states relative to the angular momentum of the
final state pion: for odd $l_\pi$ only singlet-even initial
states are possible and for even $l_\pi$ only triplet-odd.
The separation of initial spins for different parities leads
to a symmetric unpolarized cross section as stipulated by the
Barshay-Temmer theorem for pure isospin reactions \cite{bt}.
Obviously the presence of a class IV force can mix some
isospin zero component in the initial state with opposite 
spin-parity assignments: Initial spin states will
then have both parities involved. Consequently
 the cross section is no more 
exactly symmetric about $90^\circ$ \cite{cheung}. 
Presently there is an on-going experiment at TRIUMF 
\cite{e704} attempting to measure this asymmetry
in the reaction $np \rightarrow d\pi^0$.

In Ref. \cite{nst} a few CSB mechanisms were studied 
in this reaction above 350 MeV. 
As a class IV force behaves spatially much like
the spin-orbit force, in low energy scattering
 its effects vanish. In  $np \rightarrow d\pi^0$ one
can also argue for the smallness of asymmetry at threshold,
because it must be an interference of opposite parity 
amplitudes, at least $s$- and $p$-wave pions, whereas
the symmetric cross section has a squared $s$-wave term.
Therefore, understandably
Ref. \cite{nst} found quite small asymmetries at and below
400 MeV and did not aim to any details in the threshold 
region. However, it turns out that there are experimental
advantages at threshold allowing smaller relative asymmetries
to be detected than at higher energies, even at the level
of one part in a thousand. The TRIUMF experiment E704 
\cite{e704} utilizes this feature to measure the integrated
asymmetry close to threshold. Also, theoretically at threshold
there seems to be less cancellation of possible $\eta \pi$ 
mixing effects than at higher energies studied in Ref.
\cite{nst}. This paper aims now
to provide some detailed predictions for this observable in
the threshold region where the experiment is performed.

\section{Theory}
\subsection{CSB mechanisms}
A standard source of the class IV force, dominant in most
experiments, is the $np$ mass difference in pion exchange.
Taking this into account the pion-nucleon coupling becomes
\begin{equation}
H_{\pi NN}= -\frac f \mu \, [{\vec\sigma}\cdot\nabla\,
\vec\phi\cdot\vec\tau
 + \delta\, {\vec\sigma}\cdot\nabla\phi_0
 + \delta\, \vec\sigma\cdot(\vec p+\vec p ')
(\vec\tau\times\vec\phi)_0]
\label{picpl}
\end{equation}
with the small parameter $\delta=(M_n-M_p)/(M_n+M_p)=(M_n-M_p)/2M$.
Here the first term is the familiar isospin invariant interaction
and gives rise to the well known OPE potential. The initial and
final momenta $\vec p$ and $\vec p'$ operate on the nucleons and
$\nabla$ operates on the pion field assumed a plane wave. The latter
two terms give rise to the CSB potential of the form
(using the usual notations of the literature \cite{nst})
\begin{equation}
V_\delta=\delta\,\frac{f^2}{4\pi}\,\frac\mu 3 \,
\{(\vec\tau_1+\vec\tau_2)_0\, [S_{12}\, V_T(\mu r)
+\vec\sigma_1\cdot\vec\sigma_2\, V_C(\mu r)]\,-\,
6(\vec\tau_1\times\vec\tau_2)_0\, (\vec\sigma_1
\times\vec\sigma_2)\cdot\vec L\, V_{LS}(\mu r)\}.
\end{equation}
Here the first part is of class III, conserves the isospin
and also with respect to the space and spin acts similarly
to the normal OPE (i.e. has the standard
tensor and spin-spin parts). However, the latter
term changes both the spin and isospin being of
class IV, i.e. couples the two possible spins for a 
given $L=J$ partial wave. The associated set of
coupled Schr\"odinger equations is solved numerically.

In pion physics, production, scattering as well as in
the two-nucleon interaction, the coupling of the
pion and nucleon to the $\pi N$ resonance $\Delta(1232)$
is extremely important and dominates some processes
\cite{garci,aren}.
An isospin breaking effect can arise also
 here from the mass differences between the neutron and
proton as well as between different charge states of
the $\Delta$. For charged pions this gives an isovector
correction to the standard coupling as follows \cite{nst}
\begin{equation}
H_{\pi \Delta N}= -\frac {f^*} \mu \, [{\vec S}\cdot\nabla\,
\vec\phi\cdot\vec T 
 + \delta\,\vec S\cdot(\vec p+\vec p ')(\vec T\times\vec\phi)_0 ] .
\end{equation}
Here the transition spin (isospin) operator $\vec S$ ($\vec T$)
changes the spin (isospin) $\frac 1 2$ particles to those 
with $\frac 3 2$. The mass difference between consecutive
charge states of the $\Delta$ has been assumed to be the
same as for the nucleons. In the coupling (3) there is no term
corresponding to the middle term in Eq. (1). As above for the
nucleons one gets an isospin symmetry breaking transition
potential
\begin{eqnarray}
V^{tr}_\delta({\rm OPE})&=&
\displaystyle \frac{\delta ff^*}{4\pi}\,\frac\mu 3 \,\left\{T_{10}\,
[S^{\rm II}_{12}\, V_T(\mu r)+\vec S_1\cdot\vec\sigma_2\, V_C(\mu r)]\,
 \right. \nonumber  \\
&-& \left. 6(\vec T_1\times\vec \tau_2)_0\,
(\vec S_1\times\vec\sigma_2)\cdot\vec L\,
V_{LS}(\mu r)\right\} +(1\leftrightarrow 2). \label{tranpot}
\end{eqnarray}
In the tensor operator $S^{\rm II}_{12}$ 
now one spin operator has been replaced by
the corresponding transition spin operator. All terms
arise analogously to the $NN$ case, the first terms from CSB
at the nucleon vertex.

A notable feature in this isospin breaking transition potential
 is that, contrary to the case of the $NN$ interaction, 
also the first term in (\ref{tranpot}) 
(analogous to the class III term) can cause a transition 
from an isospin zero $NN$ state to an intermediate $\Delta N$
state which can directly participate in pion production.

In addition to the possibility of the isospin mixing in the 
initial $np$ state, the  pion coupling (\ref{picpl}) gives a 
possibility also for isospin breaking in the vertex
generating the final pion state. 
From the baryon point of view the middle 
term is like a coupling of an isoscalar meson, since there
is no isospin operator. This means that there is a finite
amplitude of {\it direct transition} from an initial isospin
zero state to the deuteron state (and the pion). Of course,
there is the same small parameter $\delta$ associated as in 
the isospin breaking potentials. \\

There are other possible isospin breaking mechanisms. 
Analogously with the above effective isoscalar meson 
coupling, also production of
first a true off-shell isoscalar pseudoscalar meson 
($\eta\; {\rm or}\;\eta'$)
is possible with its subsequent transformation into pion,
 because there is a nonvanishing mixing between the $\eta$ and 
$\pi$ mesons \cite{mcnamee,coon}. The coupling of pions
to nucleons via this is of the form
\begin{equation}
H^{\rm prod}_{\eta\pi}=
-\,\frac{f_\eta} \mu\,\frac{\langle\eta|H|\pi\rangle}
{\mu^2-\eta^2}\,\vec\sigma\cdot\nabla\phi_0 
\end{equation}
taking into account the two time orderings of the production
and mixing interactions. (Ref. \cite{nst} had only the main one.)
Using the mixing matrix $\langle \eta | H |\pi\rangle = -5900 
\,{\rm MeV}^2$ \cite{coon} and the $\eta NN$ coupling
$G_{\eta}^2/4\pi = 3.68$ \cite{dumb}
with $f_\eta = G_{\eta}\mu /2M$
it can easily be seen that the strength of this contribution
should be about 15 times larger than the isoscalar meson like
coupling of the pion from the $np$ mass difference in 
Eq. (\ref{picpl}).
So one would expect this to be a very important effect.
This is further enhanced by the $\eta'$ meson mixing with 
the mixing matrix element $-5500$ MeV$^2$ \cite{coon}. 
(The coupling of the $\eta'$ to the nucleon is taken to be 
the same.) 

There are great uncertainties in the $\eta NN$ and $\eta' NN$ 
coupling strengths $G_\eta$. Much 
smaller values are also quoted from pion photoproduction
\cite{tiator} and a sensitive probe for this coupling 
is desirable to clarify the situation. The above value is 
obtained in a meson exchange $NN$ potential model fit to 
elastic $NN$ scattering and is 
consistent (in the upper end) with the range 2--7 given in
various versions of the Bonn potentials \cite{elster}.
The latter also include a form factor of the monopole form
at the $\eta NN$ vertices with $\Lambda$=1500 MeV (one
potential uses even 2000 MeV). For this particular part of
isospin breaking the former value of the cutoff is adopted
as well as the Bonn cut-off 1300 MeV is also used for the pion.
It should be further noted that the above value quoted in Ref.
\cite{dumb} is given as the strength of the meson coupling,
i.e. for $|{\bf k}|=0$ rather than at the meson pole. So in 
the normal Bonn parametrization this would correspond to
$G_{\eta}^2/4\pi = 4.8$, very close to the most common value
5 in the Bonn potentials.

Another uncertainty is related to a controversy
of off-shell $\rho\omega$-meson mixing. The mixing
matrix for this is determined on-shell, since both mesons
have basically the same mass. Arguments have risen, 
based on fermion loop calculations and QCD sum rules, that
the effect of $\rho\omega$ mixing should be much smaller
than previously estimated in class IV interactions,
because of the alleged off-shell momentum dependence 
of the mixing matrix \cite{off}. However, this is contested 
in a newer analysis of the mixing matrix \cite{coon2}.
The effect of this off-shell modification, if necessary,
would be to make this contribution nearly negligible in the
$\Delta A$ of $np$ elastic scattering \cite{offeff}
causing trouble with the data at low energy \cite{knutson}. 
A similar effect could affect also $\eta \pi$ mixing
\cite{pieoff}. However, in this 
case one particle is always off-shell even in determinations 
of the mixing matrix, because the masses are so different.

In the $NN$ sector $\eta\pi$ mixing causes only a class III
force. As seen above for the pion even this kind of
coupling can produce an $NN \rightarrow \Delta N$
transition potential
\begin{eqnarray}
V^{tr}_{\eta\pi}&=&-\,\frac{f_\pi^*f_\eta}{4\pi}\,\frac\mu 3
\,\frac {\langle\eta|H|\pi\rangle}{(\eta^2-\mu^2)}\,
T_{10}\, \left\{  
S^{\rm II}_{12}\left[V_T(\mu r)-\left(\frac\eta\mu\right)^3
V_T(\eta r)\right] \right. \nonumber \\
&+&\left. \vec S_1\cdot\vec\sigma_2\left[V_C(\mu r)
-\left(\frac\eta\mu\right)^3V_C(\eta r)\right]\right\}
+(1\leftrightarrow 2)  ,
\end{eqnarray}
which can act also in isospin zero initial states.
This potential includes now all time orderings of the meson
production, absorption and mixing interactions
and looks like a pion exchange with an $\eta$-ranged
cut-off.
In this way, as a two-step process also $\eta\pi$
mixing can produce an isospin mixing effect even in
$np$ scattering \cite{austr}. Due to the rather strong 
effective coupling seen above, also this should have a significant 
effect in pion production. Of course, the pion and $\eta$
form factors are included in the potential. \\

The isospin symmetric amplitudes are calculated in a standard
way generating the important isobar configurations 
in the initial states by solving coupled Schr\"odinger 
equations (coupled also with isospin zero $np$ states)
as described e.g. in Ref. \cite{ppdpi}, the procedure dating
back basically over two decades \cite{aren2}. 
This accounts then for both the
direct production mechanism and pion rescattering through
the $\Delta$. Also pion $s$-wave rescattering 
from the second nucleon is taken into
account in production. Details of the present potentials 
can be found
in Ref. \cite{csb2}. They reproduce the height of the 
$pp\rightarrow d\pi^+$ well and $NN$ phase shifts to
an accuracy of a few degrees from threshold over the
$\Delta$ region. Here the pion cut-off is softer, but
this transition potential is in the present context 
chosen to give the right overall strength of the 
$NN \rightarrow \Delta N$ transition, whereas the CSB one
has the previously described parametrization to be
consistent with the sources of the $\eta NN$ coupling
constant.

\subsection{Amplitudes}
The asymmetry of the unpolarized cross section
arises because there are opposite 
parities interfering in the same spin states.
Close to threshold it involves just an 
interference of $s$- and $p$-wave pions, mainly of 
the CSB amplitude
$^1P_1 \rightarrow ^3P_1 \rightarrow s$ with the important $^1D_2
\rightarrow ^5S_2(\Delta N) \rightarrow p$, if CSB is constrained 
to the $NN$ sector. Due to the $P$-wave in the initial states
the CSB $\Delta N$ mixing is not expected to be very important.
However, CSB $p$-wave pions can arise from
$^3D_2 \rightarrow ^1D_2 \rightarrow p$ or from the $\Delta$
excitation process
$^3D_2 \rightarrow ^5S_2(\Delta N) \rightarrow p$ interfering with
the dominant $^3P_1 \rightarrow s$ amplitude. This process
has the advantage of producing the $\Delta N$ intermediate state
in a lower angular momentum state than the initial nucleons. 
This is similar to the isospin respecting $^1D_2
\rightarrow ^5S_2(\Delta N) \rightarrow p$ process showing
a resonant (and dominant) structure in the $\Delta$ region.

Due to angular
momentum and parity conservation, the class IV interaction between
nucleons can only connect  singlet and triplet states with
$L=J$, i.e. only tensor uncoupled states. However, 
in an expanded baryon space also tensor
coupled isospin zero states can experience isospin breaking
transition to $\Delta N$ intermediate states \cite{csbdeut}. 
This brings in e.g. the transition chain $^3D_1 \rightarrow
^3S_1(\Delta N) \rightarrow p$ for $p$-wave pion
production, analogous to the dominant $^1D_2 \rightarrow
^5S_2(\Delta N) \rightarrow p$ and possibly a resonant
structure at the $\Delta$ energies. Also $^3S_1 \rightarrow
^3S_1(\Delta N) \rightarrow p$ is possible.

 Overall, all the processes 
considered above lead to indistinguishable
final states and must be added coherently in the
amplitudes.

\section{Results and conclusion}
In Table \ref{amptable} complex contributions to CSB partial
wave amplitudes from various components already discussed 
above are presented separately at the laboratory energy
279.5 MeV. The $np$ mass difference in pionic potentials acts
mainly in the $NN$ sector in this energy region. The class IV
force has no effect in the tensor coupled $np$ states as
shown above, whereas isobar configuration mixing gives 
a small contribution also there. However, the effect of
these configurations is suppressed by nearly an order of 
magnitude as compared with the $NN$ contribution. The
effects at this energy are nearly the same in $s$-wave
and $p$-wave production (from $^3D_2$) even at this low 
energy. At higher energies 
$p$ waves gain in importance in proportion to 
the pion momentum. Clearly at this energy the pionic 
$d$ waves and higher can
be omitted: they are suppressed by an order of magnitude.

\begin{table}[tb]
\caption{Relative importance of contributions to CSB amplitudes
at the laboratory energy 279.5 MeV
 from the isoscalar meson like pion production vertex, isospin
symmetry breaking pionic $NN$ potentials due to the $np$ mass
difference in the $NN$ sector and
in the extended two-baryon space, $\eta$ and $\eta'$
production followed by transformation into a $\pi^0$ and potentials
involving $\eta\pi$ or $\eta'\pi$ mixing.}

\begin{tabular}{cccccc}
 & & & & & \\
Amplitude  & $\pi$ vertex &
$\pi$ pot. ($NN$)& $\pi$ pot. (tot.) &
$\eta$ vertex  & $\eta\pi$ potential \\ \hline
$^1P_1\rightarrow s$ & (-0.45,0.24) & (0.25,0.60) &
 (0.10,0.66)  & (-9.04,4.78) & (-1.24,0.10) \\
$^3S_1\rightarrow p$ & (-0.21,-0.07)& 0           &
 (0.02,0.01)  & (-4.09,-1.39) & (0.42,0.09)  \\
$^3D_1\rightarrow p$ & (-0.35,0.13) & 0           &
 (0.01,-0.00) & (-7.06,2.54) & (0.14,-0.02) \\
$^3D_2\rightarrow p$ & (0.24,0.10)  & (0.05,-0.58)&
 (-0.04,-0.69)& (4.85,2.07)  & (-2.21,-1.29)\\
$^1P_1\rightarrow d$ & (-0.02,0.01) & (0.00,-0.01)&
 (0.01,-0.01) & (-0.32,0.17) & (0.02,0.00)  \\
$^1F_3\rightarrow d$ & (-0.01,0.00) & (0.00,-0.02)&
 (0.00,-0.02) & (-0.30,0.00) & (0.07,0.00) \\ \hline \\

\end{tabular}
\label{amptable}
\end{table}

The isoscalar meson coupling like production
vertex (from the middle term in Eq. (\ref{picpl})) 
gives a contribution  of the
same order of magnitude, but appears with the same strength 
also in the tensor coupled $^3S_1$ and $^3D_1$ initial
states. There its effect is much more than the coupled
isospin breaking channels - nearly all comes from the
isospin breaking in the production vertex.

As anticipated earlier, the $\eta\pi$ mixing in the production 
vertex is about twenty times stronger than for the pion.
As the $\eta\pi$ mixing potential appears effectively only in 
the $\Delta$ chains with the pion and $\eta$ cutting off each 
other, the relative role of the production vertex is even enhanced.
In the threshold region this single mechanism rises above 
the others in importance. The isobar effect arising from
$\eta\pi$ mixing is overall still stronger than the class IV
mixing in the $NN$ sector from OPE. In these comparisons the
absolute normalization and dimension are immaterial, but
in relating different $J$ values it should be known that
these are reduced matrix elements of the two-nucleon system
in the sense of Ref. \cite{desh} with the $3j$-normalization.
Furthermore, it is of interest to note that 
 at this energy the sizes of the isospin conserving
$s$- and $p$-wave (from $^1D_2$) amplitudes just happen to be a
hundred times those of the $\eta$-vertex column ($p$ wave from
$^3D_2$).  \\

The quantity of experimental interest here is the integrated
forward-backward asymmetry divided by the total reaction
cross section 
\begin{equation}
A_{fb} \equiv \int [\sigma(\theta) - \sigma (\pi - \theta)]
d\Omega \, / \int \sigma(\theta) d\Omega  .
\end{equation}
Here the angle is the CM angle between the deuteron and
incident neutron directions. In Table \ref{fbtable} this is
given in per cent (i.e. it is multiplied by 100) for a range
of energies in the neighbourhood of threshold.

\begin{table}[tb]
\caption{Contributions to the integrated forward-backward asymmetry
(\%) from the isoscalar meson like pion production vertex, isospin
symmetry breaking pionic $NN$ potentials in the $NN$ sector and
in the extended two-baryon space, $\eta$ and $\eta'$
production followed by transformation into a $\pi^0$ and potentials
involving $\eta\pi$ or $\eta'\pi$ mixing. The total should be the sum 
of $\pi$ vertex, $\pi$ pot. (tot.), $\eta$ vertex and $\eta\pi$
potential. For comparison, the anticipated precision of the
experiment E704 \protect\cite{e704} is 0.12 \% .}

\begin{tabular}{ccccccc}
 & & & & & \\
$E_{\rm lab}$ (MeV) & $\eta =q/m_\pi^0$ & $\pi$ vertex &
$\pi$ pot. ($NN$)& $\pi$ potential &
$\eta$ vertex  & $\eta\pi$ potential \\ \hline
278  & 0.138 & -0.008 & 0.026 & 0.020 & -0.156 & -0.117 \\
279.5 & 0.170 & -0.009 & 0.032 & 0.024 & -0.183 & -0.141 \\
281 & 0.197 & -0.010 & 0.036 & 0.028 & -0.205 & -0.161 \\
285 & 0.255 & -0.012 & 0.045 & 0.035 & -0.247 & -0.201 \\
290 & 0.314 & -0.014 & 0.054 & 0.041 & -0.276 & -0.236 \\
300 & 0.408 & -0.015 & 0.064 & 0.049 & -0.292 & -0.283 \\
320 & 0.555 & -0.013 & 0.073 & 0.056 & -0.260 & -0.337 \\
350 & 0.730 & -0.009 & 0.077 & 0.057 & -0.176 & -0.382 \\ \hline \\

\end{tabular}
\label{fbtable}
\end{table}

The third column shows the contribution from the CSB
isoscalar pion  coupling at the production vertex. 
This is rather a minor contribution and is, in fact,
 more than cancelled by isospin breaking 
pion potentials (columns 4 and 5), mainly OPE in the
nucleon sector at these energies, but also some amount
from transitions into $\Delta N$ intermediate states.
Both of these are dwarfed by the effects of $\eta\pi$
mixing. The column denoted by "$\eta$ vertex" gives again the 
contribution from the CSB production vertex: a generated off-shell
$\eta$ meson changes into the final state on-shell pion. This 
is the dominant contribution but not at all as massively as
expected from the sizes of the amplitudes.
Comparably important comes here the isospin breaking 
$\eta\pi$ transition
potential and the importance of the latter increases
with increasing energy. In the $\Delta$ region it is the 
largest individual contribution \cite{nst}. This happens, 
because both production vertex effects turn back down at
the highest energies shown. In fact, in the neighbourhood
of $E_{\rm lab}\approx 400$ MeV they even change sign
leading to significant cancellation of the $\eta\pi$ mixing 
effects in the $\Delta$ region as was found in Ref. \cite{nst}. \\

Preliminary calculations indicate that also the $\rho$
and $\rho\omega$-mixing effects as well as the electromagnetic 
interaction are significantly smaller
than $\eta\pi$ mixing. These are 
in the same order as the pion effects. Therefore,
as a summary it seems that
CSB threshold production is strongly dominated by $\eta\pi$
mixing suggesting CSB measurements as an effective tool 
to study this phenomenon and constrain the $\eta NN$ coupling.

\end{document}